\documentclass[graybox,natbib,nosecnum]{svmult}
\bibpunct{(}{)}{;}{a}{}{,} % suppress commas between author-names and year

\pdfoutput=1   %forces use of pdflatex. Disable if you prefer to use .eps or .ps figures.
\usepackage{mathptmx}       % selects Times Roman as basic font
\usepackage{helvet}         % selects Helvetica as sans-serif font
\usepackage{courier}        % selects Courier as typewriter font
\usepackage{type1cm}        % activate if the above 3 fonts are
\usepackage{makeidx}         % allows index generation
\usepackage{graphicx}        % standard LaTeX graphics tool
\usepackage{multicol}        % used for the two-column index
\usepackage[bottom]{footmisc}% places footnotes at page bottom
\usepackage[normalem]{ulem}	% for strike-through of text with \sout{}  
\usepackage{hyperref}  %for hyperlinks
\usepackage{eurosym} % For euro symbol \euro -> €
\usepackage{amssymb}  % For math symbols, e.g. \gtrsim

\usepackage{soul}   % for high-lighting of text
\makeindex             % used for the subject index
\begin{document}

\title*{Future Astrometric Space Missions for Exoplanet Science}
\author{Markus Janson, Alexis Brandeker, Celine Boehm, and Alberto Krone Martins}
% Use 
\authorrunning{M. Janson, A. Brandeker, C. Boehm, and A. Krone Martins} 
%for an abbreviated version of
\institute{
Markus Janson \at Department of Astronomy, Stockholm University, AlbaNova University Center, SE-10691 Stockholm, Sweden, \email{markus.janson@astro.su.se}
\and 
Alexis Brandeker \at Department of Astronomy, Stockholm University, AlbaNova University Center, SE-10691 Stockholm, Sweden, \email{alexis@astro.su.se}
\and 
Celine Boehm \at Institute for Particle Physics Phenomenology, Department of Physics, Durham University, DH1 3LE Durham, United Kingdom, \email{c.m.boehm@durham.ac.uk}
\and 
Alberto Krone Martins \at CENTRA/SIM, Faculdade de Ciencias, Universidade de Lisboa, Ed. C8, Campo Grande, P-1749-016 Lisboa, Portugal, \email{algol@sim.ul.pt}
}
\maketitle

\abstract{High-precision astrometry at the sub-$\mu$as-level opens up a window to study Earth-like planets in the habitable zones of Sun-like stars, and to determine their masses. It thus promises to play an important role in exoplanet science in the future. However, such precision can only be acquired from space, and requires dedicated instrumentation for a sufficient astrometric calibration. Here we present a series of concepts designed for handling this task. STARE is a small satellite concept dedicated to finding planets in the very nearest stellar systems, which offers a low-cost option toward the study of habitable planets. The NEAT concept is a set of two formation-flying satellites with the aim to survey the 200 nearest Sun-like stars for Earths in the habitable zone. Finally, THEIA is a proposal for an ESA M-class mission, with a single-unit telescope designed for both dark matter studies as well as a survey for habitable Earth-like planets among the 50 nearest stars. The concepts illustrate various possible paths and strategies for achieving exquisite astrometric performance, and thereby addressing key scientific questions regarding the distribution of habitability and life in the universe.}

\section{Introduction }

In some sense, astrometry has always been the future of exoplanet research. Or in more precise terms, it is a technique whose future prospects for revolutionising the field is obvious and has been realized since early on, but at the same time, the challenges of the technique have continually hampered its utility in the present. There is good reason to expect that the final results of \textit{Gaia}, which is discussed elsewhere in this book, will drastically change this situation, making astrometry a frontline field in exoplanetary science. However, the historical yield of astrometric exoplanetary science pre-\textit{Gaia} has been rather modest:

To date, there have been no primary detections (i.e., verified detections of planets that had never been previously observed) of exoplanets with astrometry. This is not for a lack of attempts. In fact, the first claimed discovery of an exoplanet candidate around Barnard's star in 1963 was based on astrometric data \citep{vandekamp1963}. However, several subsequent studies with similar or better precision have yielded null results, thus firmly excluding the existence of such an exoplanet \citep[e.g.][]{gatewood1973,benedict1999}. In more recent history, an astrometric planet candidate was suggested to orbit the nearby low-mass star VB~10 \citep{pravdo2009}, but again, follow-up data by e.g. \citet{bean2010} could unambiguously discard this hypothesis. These cases illustrate the critical need to have a good understanding of the systematics affecting the astrometric performance, which involve effects arising from the optical system, as well as from the Earth's atmosphere if the observations are acquired from the ground. Some useful secondary astrometric detections or upper limits have nonetheless been acquired up to now, mostly using the space-based telescopes \textit{Hipparcos} and \textit{Hubble} \citep[e.g.][]{benedict2010,mcarthur2010,reffert2011}.

Nonetheless, as we allude to above, the future of exoplanet astrometry is highly promising, given the capabilities of the technique as well as the instrumental development necessary to recover astrometric planet signals. In the remainder of this introduction, we will discuss the need for and advantages of astrometry for exoplanet science, and in the three sections that follow, we will present different concepts that are in development for realising such science. 

There are a number of fundamental reasons for why astrometry is useful for understanding exoplanets. Firstly, while radial velocity alone only yields a minimum companion mass due to an ambiguity between mass and inclination, astrometry breaks this ambiguity, providing the actual mass and inclination. This avoids false positives in the form of e.g. brown dwarf or low-mass stellar companions on nearly face-on orbits, imitating the signal of a more highly inclined planet. It also means that astrometry is sensitive to planets of any orbital orientation, as opposed to radial velocity which is insensitive to face-on orbits (and even more so for transits, which are exclusively constrained to nearly edge-on orbits). Microlensing can yield an actual planet mass if the host star can be identified and characterised, but is limited to rare and random events. Hence, astrometry is the only existing method that can be routinely used for mass determinations of arbitrary targets.

In contrast to radial velocity, the amplitude of the astrometric signal increases with increasing orbital semi-major axis. This makes it highly complementary to direct imaging, which has a similar bias toward larger orbits. Thus, there is considerable promise for a beneficial synergy between astrometry and direct imaging in the future, in the same way that a synergy between transits and radial velocity exists today. Direct imaging can yield a wealth of useful information about planets such as temperature, size, and atmospheric composition, but cannot yield a mass. Furthermore, direct imaging is sensitive to the orbital phase, which means that a blind imaging survey for planets will tend to miss some targets that should nominally be detectable, because the planet resides in an unfortunate phase at a small angular separation from the star. Astrometry can help with both of these aspects, where an astrometric survey of an input sample can identify all targets with interesting planets, provide ephemerides for the purpose of ideal scheduling of direct imaging follow-up, and provide masses for all such targets.

An additional advantage of astrometry relative to radial velocity is that it is relatively insensitive to stellar noise due to activity and spots \citep{lagrange2011}. Hence, it is less susceptible to false positive detections from such noise sources, which is a relatively common issue in radial velocity surveys \citep[e.g.][]{hatzes2013,robertson2014}. This also means that astrometry has considerable advantages regarding studies of particularly active stars, such as low-mass stars, or young Sun-like stars. Surveying young stars in addition to older ones allows to investigate the formation and evolution of planets to an extent that is otherwise not possible. Young systems are also another area in which astrometry offers fruitful synergy with direct imaging -- young planets are hot, and therefore emit considerable amounts of thermal radiation, making them easier to detect in imaging.

One of the most important tasks in exoplanet research within the next few decades is to establish a complete census of Earth-like planets in the solar neighbourhood. This will be a sample that can be well characterised, and that can form the basis for well-posed statistical questions regarding the frequency of habitability and life beyond the solar system. From the previous reasoning in this chapter, it follows that astrometry is the only method that can accomplish this -- radial velocity has a bias against low-inclination orbits, the vast majority of systems do not undergo transits or microlensing events, and direct imaging is phase dependent. This, in combination with technological developments that make high-precision astrometry increasingly feasible, motivates the development of the astrometric exoplanet detection concepts described in the following sections. 

The concepts discussed represent different technical approaches to reaching an astrometric precision sufficient for studying Earth-like exoplanets, which is the next natural step after the expected legacy of \textit{Gaia}. First, we will discuss the small-scale specialized concept STARE, and then two concepts for larger scale projects with a broader scientific scope called NEAT and THEIA. The latter is closely related to the STEP concept, which is not described in detail here, but represents another potential platform for high-precision astrometry.

\section{STARE}

The great distances between stars constitutes one of the main challenges for high-precision astrometry: Firstly, it means that bright stars are scarce in the sky, so if we wish to relate the motion of a target star to a reference frame of other stars, we typically need to do so over considerable angular separations. For such so-called `wide-angle' astrometry, we are subject to beam walk effects that impact the achievable precision, and furthermore, covering a very wide field of view requires large detectors, which are extremely costly. Secondly, the astrometric signature $\alpha$ scales inversely with distance $d$, so even at a distance of 10~pc, which is small by galactic standards, the astrometric signature of an Earth-equivalent planet around a Sun-like star is a minuscule 0.3~$\mu$as.

As it happens, both of these issues can be mitigated by focusing on the very most nearby stellar system beyond our own: the $\alpha$~Centauri system. $\alpha$ Centauri consists in principle of three stars, but here we will focus on the two that are the most Sun-like: $\alpha$~Cen A and B. With a distance of only 1.3~pc, an Earth-equivalent planet around either star would impose an astrometric signature of $\sim$2.5$\mu$as, thus greatly alleviating the required precision relative to a 10~pc sample. In addition, with a semi-major axis of 17.6~AU, the two components are separated by several arcseconds on the sky ($\sim$4--10$^{\prime \prime}$ until the next periastron passage in 2035), and thus widely separated relative to the resolution limit even of a modest-sized telescope. Thus, if the A and B stars are used as astrometric references for each other, only narrow-angle astrometry is necessary, which benefits the achievable precision. Furthermore, being so nearby, the system is extremely bright ($V=0$~mag), and thus a high S/N can be built up even over extremely short timeframes. 

STARE is a concept for achieving the sensitivity necessary for detecting Earth-like planets in the classical habitable zone (CHZ) at a low cost, by using novel technology on a small-aperture (12.5~cm diameter) telescope and focusing almost exclusively on the $\alpha$~Cen system. The required precision will thus be reached by collecting large amounts of photons, helped by the near-continuous monitoring of one of the brightest targets in the sky, and by minimizing and calibrating for systematic effects. The critical aspects for the latter purpose are thermal stability, and the radiation environment.

The tube for the STARE telescope will be constructed in carbon fibre with its most stable axis along the optical axis, and the optical components will be made in Zerodur. The pre-camera part of the telescope will be kept thermally stable to a $\sim$1K-level precision. This enables a good optical stability, but still, any residual instability can cause the image scale factor of the telescope to change, which will change the apparent separation of the A/B binary and thus impact the astrometric precision. Hence, it is desirable to independently monitor the scale factor. In STARE, this is done by splitting up the wavefront of the incoming light prior to the focusing optics, as will be described in more detail below. The detector uses a scientific CMOS (sCMOS) chip of 2k by 2k pixels, which is read out in global shutter mode at a 50 Hz rate (i.e., 20 ms per frame). The detector is strongly oversampled (a factor 20 relative to the FWHM) which partly serves to average out flat field-related errors, and along with the high frame rate, it keeps the targets comfortably within the linear response regime of the detector.

\begin{figure}
\includegraphics[scale=0.2]{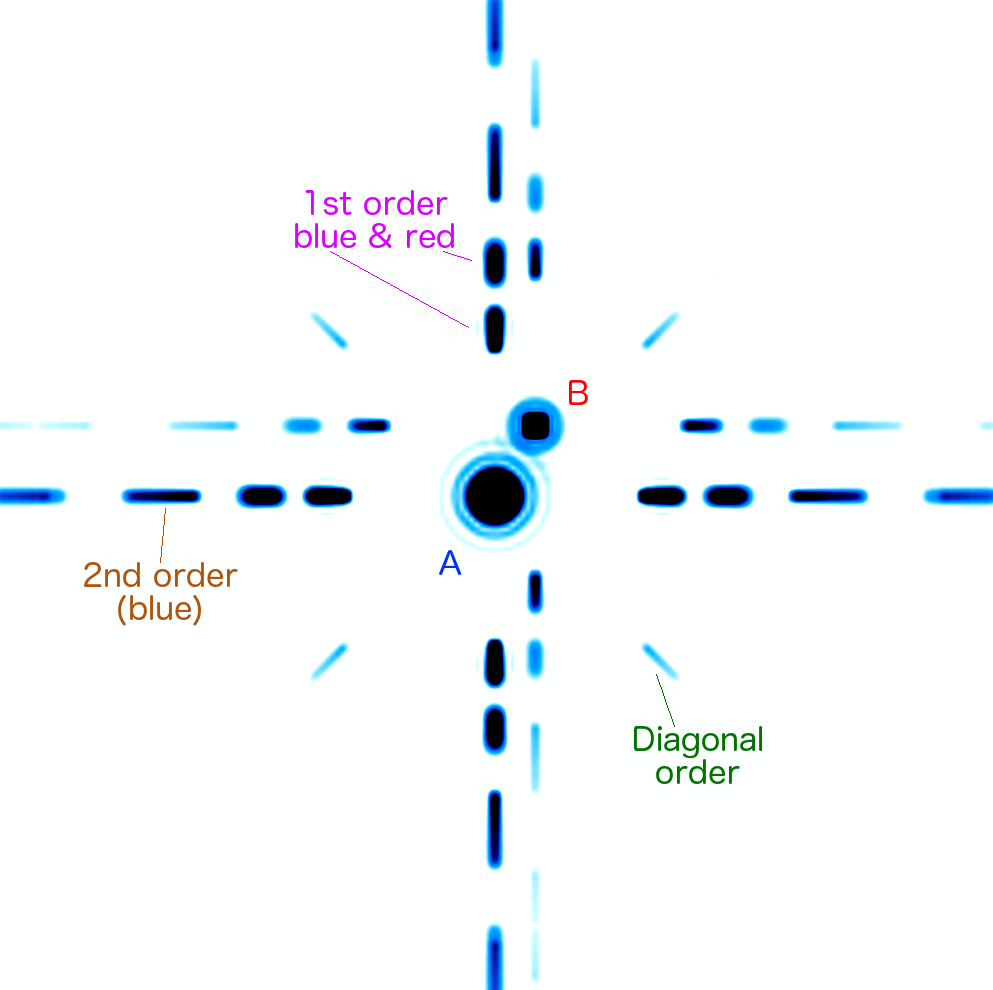}
\caption{Simulated point spread function of the STARE concept. `A' and `B' denote the two stellar components of the binary. In addition to the main modes of conventional PSFs, each star has several higher-order PSF satellite features. Each order is separated into two distinct colours by wedge filters.}
\label{f:starepsf}  
\end{figure}

The wavefront splitting mentioned above is done with a form of transmissive phase grating, which produces dispersive first- and second-order PSF features at fixed angular separations from the non-dispersive zero-order features of the two stars (see Fig. \ref{f:starepsf}). In order to achieve sharp edges of the dispersive features in the image plane, wedge filters are used to produce a transmission profile which is close to 1 between 450 and 550 nm and between 650 and 750 nm, and zero elsewhere. In other words, each spectrum is split up into a blue and a red spot. This gives good astrometric information thanks to the sharp edges, and additionally, provides colour information that would otherwise be unavailable. The filters are placed in a finely temperature controlled environment ($\sim$10\,mK precision) along with the detector, in order to maintain stability of the transmission edges.

The current conceptual implementation of STARE has it placed in a sun-synchronous dusk-dawnLow Earth Orbit (LEO) at a height of $\sim$650 km. This means that solar insolation will be relatively constant across an orbit and across the year, again to benefit thermal stability. The telescope will look at $\alpha$~Cen at any point during which it is visible, which is any time the satellite doesn't pass behind Earth. During these times ($\sim$20~min per $\sim$90~min orbit), it will look at photometric and astrometric calibration sources. It will also look at the 61 Cygni system on some of these occasions, which is the second nearest wide binary of Sun-like stars, and thus also well suited for astrometric planet searches. We estimate that a sensitivity down to 6~$M_{\rm Earth}$ super-Earths can be detected in the CHZs of 61 Cyg A and B, while a sensitivity below 1~$M_{\rm Earth}$ in the CHZs of $\alpha$~Cen A and B.

Through its intensive monitoring of $\alpha$~Cen, STARE will not only yield high astrometric precision, but also high photometric precision, at an extremely high cadence. This enables rich ancillary science concerning activity and asteroseismology of the stars themselves. The monitoring of activity also feeds back to the astrometric precision, since variability in the colours of the stars will affect the scale factor calibration with the dispersive PSF features discussed above. The fact that colour information is given directly by the spots is highly beneficial in this regard. Furthermore, the quasi-continuous photometric monitoring of the $\alpha$~Cen system also means that if anything Earth-sized transits either star \citep[e.g.][]{demory2015}, it would be discovered by the mission.

The most important factors that influence the performance of STARE are thermal stability (spatially and temporally) and the radiation environment, which is rather fierce in LEO due to the South Atlantic Anomaly (SAA), and which mostly affects the detector. In this context, sCMOS detectors are highly beneficial since they have a small cross-section to cosmic ray hits, and since they do not have strong memory effects, unlike CCD detectors. Our simulations thus far indicate that it will be possible to self-calibrate all the relevant aspects of the detector (flat field and non-homogeneous sensitivity across the elements) on a timescale of 40 minutes, i.e., between each SAA passage.

Because of the small size and relatively modest complexity of STARE, it fits within the scope of satellite programs by individual national space agencies, or small/fast programs of larger agencies. It can therefore effectively function as a pathfinder for larger-scale missions. $\alpha$~Cen being the easiest target for astrometric planet searches and subsequent follow-up means not only that a mission such as STARE may be very fruitful, but also that it provides a stepping stone toward broader surveys that would be necessary to attain a statistical census of the prevalence of habitable Earth-like planets in the galaxy and the universe.

\section{NEAT}

To effectively survey the nearest 200 solar-type (F, G, and K) stars for Earth-like planets in the CHZ, several challenges have to be met: A sufficient number of bright enough background sources need to be available as astrometric references in order to properly measure narrow-angle astrometry. At the same time, systematic noise due to variations in the optical path and the detector needs to be controlled. Considerations have led to essentially two solutions, one that is the NEAT concept presented here \citep{mal12,mal14a,mal14b}, and the other that is the Theia concept that will be presented in the next section.

NEAT (Nearby Earth Astrometric Telescope) was proposed in response to the 2010 European Space Agency Cosmic Vision call for a medium-sized mission (estimated 530\,M\euro\ mission cost including launch), and a reduced concept called micro-NEAT ($\mu$NEAT) was proposed in response to the 2012 ESA call for a Small Mission (70\,M\euro\ including launch). The NEAT idea is to use a single mirror off-axis parabola telescope (of diameter 1 metre) to avoid any errors resulting from the different foot-print of stellar beams emitted from different directions on any secondary or tertiary mirror. The mirror defaults at different physical locations on any secondary and tertiary mirror would introduce prohibitive astrometric errors. Using a single mirror telescope solves this problem with the trade-off that a long focal length (40\,m) is required for an adequate angular resolution. A solution to handle such a long separation between the mirror and the focal planes is to formation fly two spacecraft, one mirror carrier and one focal plane carrier.

The aberration-controlled field of view of 0.6$^{\circ}$ allows 6--9 stars with magnitude $R \le 11$\,mag to be used as references. The astrometric measurement does not rely on an inaccessible mechanical stability of the instrument, nor of inter- and intra-pixel QE responses of the CCD detector, but on an interferometric calibration of the position of 8 small movable CCDs located around a fixed central CCD. The differential astrometric measurements are done by monitoring Young's fringes originating from telescope metrology fibres located on the primary mirror. These fringes sweep the focal plane pixels several times a second, providing pixel relative positions and a quantum efficiency map every minute, during the entire 5\,yr mission. To achieve the end-of-mission goal with a 1\,$\sigma$ noise floor at 0.05\,$\mu as$, the performance of the metrology system is key; knowledge of the pixel positions at the level of $5 \times 10^{-6}$\,pixel is required.

\begin{figure}
\includegraphics[width=\hsize]{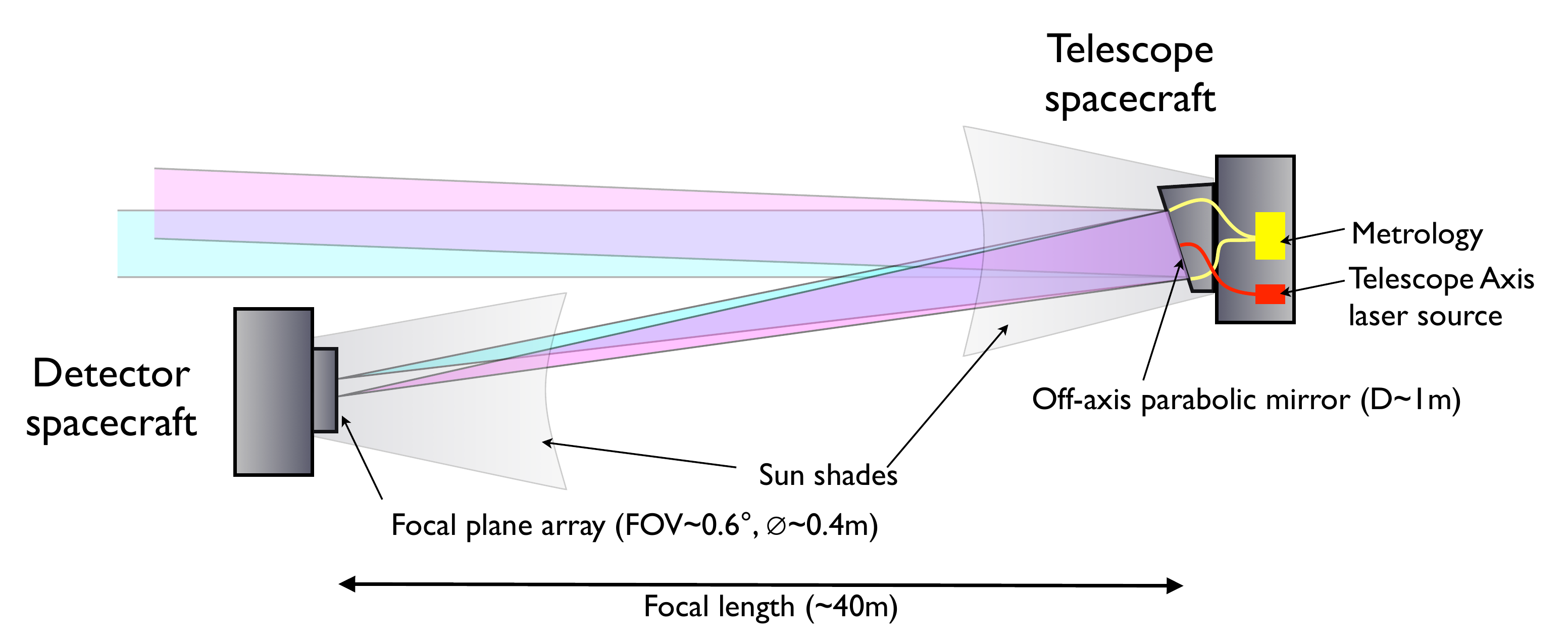}
\caption{The NEAT concept. $\mu$NEAT works in a similar way, but with a 0.3\,m mirror diameter, 12\,m focal length, reduced performance, and
a reduced price tag.}
\label{f:neat}  
\end{figure}

The metrology system greatly relaxes the requirement on the accuracy of the placement of the 8 movable CCDs. While knowledge of their positioning is required at the $5 \times 10^{-5}$\,pixel level, the actual placement only requires the point-spread functions of the 8 reference stars to be well positioned on each corresponding $512\times512$ CCD, i.e.\ a few tens of pixels placement accuracy is sufficient.

Another key technology for NEAT is formation flying. The long focal length, the metrology system, and a tip-tilt mirror help to relax the required relative positioning accuracy of the spacecraft, which are still $\le 2$\,mm for NEAT and $\le 1$\,cm for $\mu$NEAT. Consumables have to allow for $\sim$20 daily re-configurations. The PRISMA technology demonstration mission \citep{bod09} consisted of two formation flying space craft. In-flight experiments to simulate $\mu$NEAT-like manoeuvres demonstrated the capabilities to reach a 1\,$\sigma$ control stability of 4\,cm \citep{del13}. This was achieved by using off-the-shelf flight-proven hardware not specifically designed for high accuracy, including 1\,N hydrazine thrusters with a 0.7 mm/s minimum impulse bit. The factors limiting accuracy were the relatively coarse thrusters and limited-accuracy metrology combined with the stresses from a LEO, all that would be improved by several orders of magnitude by using mN ion thrusters and an active metrology system (e.g.\ LIDAR) at the planned L2 orbit location of NEAT/$\mu$NEAT.

The observational strategy for NEAT would be to visit each of $\sim$200 systems about 50 times during the 5\,yr mission, measuring the relative positions between the target and the reference stars with sub-$\mu$as accuracy for each visit. This would be sufficient to uniquely resolve a 3-planet system, including planets in the CHZ. If more planets are present, more astrometric or RV measurements would be needed to disentangle the planetary signals. Numerical simulations and a double-blind test campaign showed that the performance is in line with expectations \citep{tra10}, in particular that the detection of Earth-like planets in the CHZ were not impaired by the presence of long-period giant planets.

The ambitions of $\mu$NEAT are reduced compared to NEAT in order to fit into the ESA small mission envelope. Instead of aiming at Earth-like planets in the CHZ around 200 nearby systems, $\mu$NEAT, like STARE, is designed to be sensitive to such planets only in the $\alpha$\,Cen A \& B system. But unlike STARE, $\mu$NEAT would additionally be able to detect mini-Saturns ($\gtrsim 50$\,M$_{\oplus}$) for the 200 nearest stars and Neptunes/super-Earths ($\gtrsim 10$\,M$_{\oplus}$) for a subset of 25 stars. 

The relaxed requirement of $\mu$NEAT, 0.35\,$\mu$as end-of-mission (3\,yr) noise floor, simplifies the design significantly. The field of view is still 0.6$^{\circ}$, but the photon noise and image scale are allowed to increase, reducing the mirror diameter to 30\,cm and the focal length to 12\,m. This makes it possible to cover the focal plane array with 9 fixed $4096\times 4096$ CCDs instead of having the more complex mechanism with moving CCDs. Furthermore, the constraints on the metrology is also relaxed to the level of $4\times10^{-5}$\,pixels, a performance already achieved in the laboratory. PRISMA could be used as a platform with very minor modifications, making use of proven flight hardware and saving development cost.

$\mu$NEAT is difficult to scale down in cost, due to some key components, like the metrology system and the existing PRISMA platform, having a fixed cost. At the same time, the increase in cost for the next ambition level, represented by NEAT, is quite steep, as a new platform would have to be developed. One way to reduce the cost for the exoplanet science is to time share with other science cases that can favourably employ the instrument. This is the approach taken by Theia, described next.

\section{Theia} 

The Theia concept is the successor of the NEAT proposal, where the formation flight has been replaced by a single satellite.  The mission primarily aims to characterise the local dark matter distribution using relative astrometry to infer the dark matter  properties, but about 15$\%$ of the mission time is dedicated to the detection of Earth-like exoplanets in the Classical Habitable Zone of our 50 nearest star systems.  To maximize its impact in the community, \textit{Theia} will work as an open observatory for about 15$\%$ of the mission time. Calls for science opportunity will be opened a few years before the mission launch and scientists from the astronomy, astrophysics and cosmology communities will be invited to submit proposals. The science cases complementary to the \textit{Theia}'s primary interest (dark matter, exoplanets and compact objects) will be scrutinised, and the most promising will be selected. 

The level of precision required for the exoplanets science case (i.e. $\sim 0.3 \mu as$) exceeds the required precision for the dark matter science case by a factor 10-100. As a result, \textit{Theia} has been designed with the exoplanets science case in mind.

% The mission is foreseen to last 4.5 years (4 years observations $+$ 6 months commissioning) and most of the observations will take place at the beginning and the end of the mission (though some observations will also be performed in the meantime). 
To achieve this very challenging sub micro-arcsecond astrometry, \textit{Theia}'s strategy is to point and stare in directions of primary interest for \textit{Theia}'s science cases (dwarf galaxies, hyper velocity stars, the galactic disk and exoplanet hosts) and to minimise systematics. Thus the Theia Payload Module (PLM) was designed to be simple, being  composed of four main subsystems: a long focal length, diffraction limited, telescope; a camera nyquist sampling the PSF; focal plane array metrology to monitor FPA stability; telescope metrology to monitor the telescope structure. The mission is foreseen to last 4.5 years (including 6 months of commissioning). 

\begin{figure*}[t]
  \centering
 \includegraphics[width=0.8\hsize]{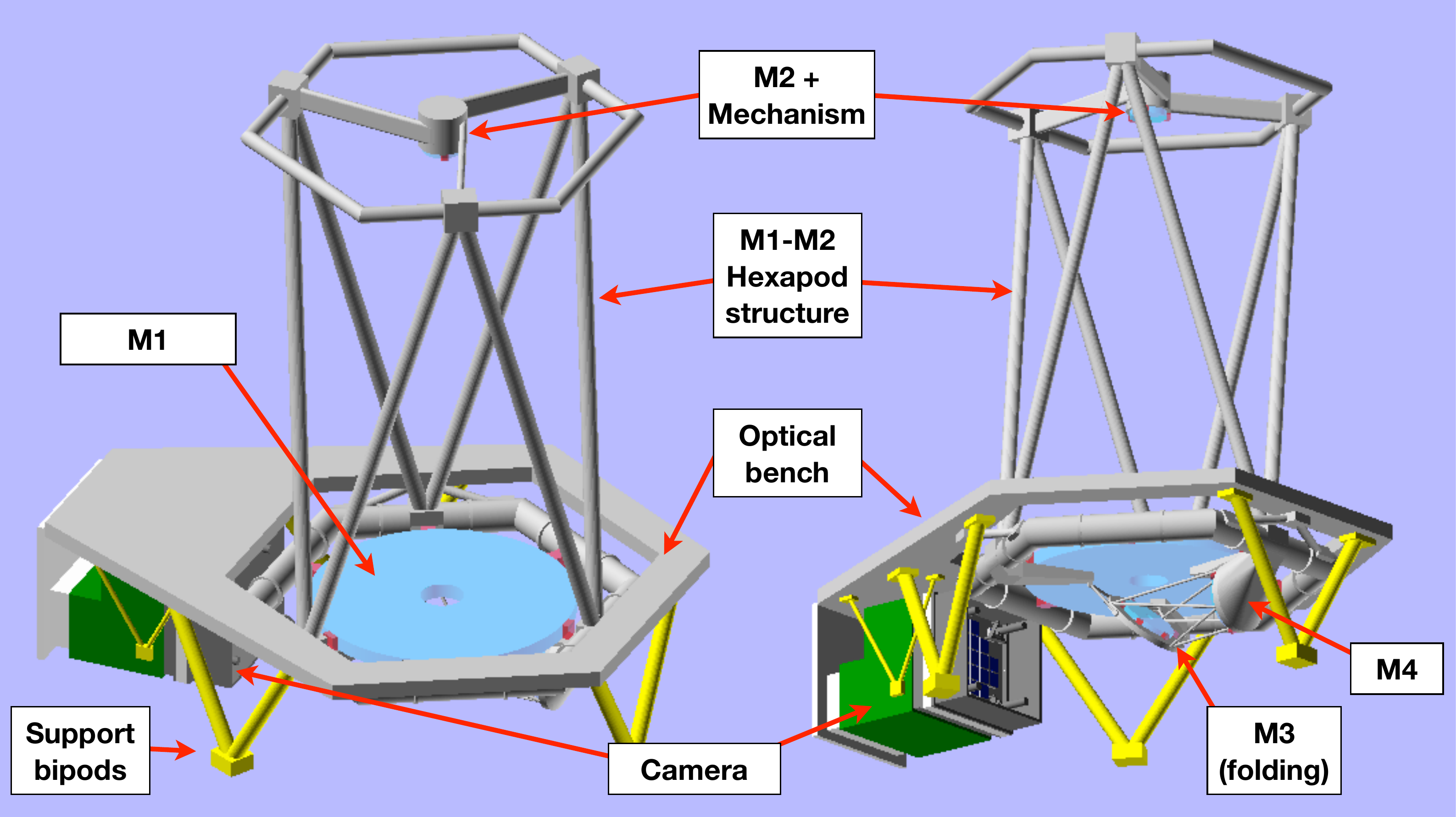}  
 \caption{Conceptual layout of the \textit{Theia} Payload Module. Volume is estimated in $1.6\times1.9\times2.2$m$^3$.}
  \label{fig:theiam5-plmconcept}
\end{figure*}

The \textit{Theia} instrument is based on a 0.8 m Three Mirror Anastigmatic (TMA) telescope with a single focal plane (see Fig.~\ref{fig:theiam5-plmconcept}), 
covering a $0.5^\circ$ Field-of-View (FoV) with a mosaic of detectors operating in the visible domain. The spacecraft will be injected into a large Lissajous orbit at L2 by an Ariane 6.2 launcher on late 2020's or early's 2030's. At L2 the absence of gravity gradients and the environmental conditions (thermal variations, lower total ionizing doses) are ideal to perform astrometric measurements. To avoid parasitic light from the Sun onto the telescope and the detector, the spacecraft will be equipped with baffles that protect the instrument from Sun light at angles larger than $\pm45^\circ$ in the Sun direction. Additionally, a stable thermal concept including a thermal shield was developed.

One of the greatest challenges is to reduce and monitor all the effects that can impact the determination of the relative positions of the point spread functions to the level of 1 part per 200 000. This requires to control i) the photon noise, which can be either dominated by the target or by the reference stars, ii)  the geometrical stability of the Focal Plane Array (FPA), iii) the stability of the optical aberrations, iv) the variation of the detector quantum efficiency between pixels. 
%As the typical apparent size of an unresolved star for such a telescope corresponds to 0.2~arcseconds, the challenge for the collaboration is to reduce all the effects that can impact the determination of the relative positions of the point spread function to the level of 1 part per 200 000. This requires to control i) the photon noise, which can be either dominated by the target or by the reference stars, ii)  the geometrical stability of the Focal Plane Array (FPA), iii) the stability of the optical aberrations, iv) the variation of the detector quantum efficiency between pixels. 

In the quiet environment of L2, the main source of instrumental variation between measurements is the thermal expansion due to the variation of the Solar Aspect Angle (SAA). To minimize astrometric shifts, \textit{Theia} adopts very low CTE materials with $10^{-6} \ \mbox{K}^{-1}$, as SiC, Si3N4 or CFRP for the structure and optical bench, Zerodur or ULE for optical components, and the mission pointings are also optimized.

To control optical aberrations up to third order, we have chosen an on-axis optical design with a single folding mirror, M3. This tertiary mirror actually has a hole in the middle for the light to reach the FPA after reflecting on M4. As a result, a fraction of the FoV (essentially a disk of diameter $\sim 0.06^\circ$) is lost in the center of the field but this does not have a strong impact on any \textit{Theia} science case. 

The FPA consists of 24 detectors arranged in a quasi-circular geometry, each detector comprising $4096(H)\times 4132(V)$ pixels of $\sim10 \mu m$. At the border of the FPA, there are four $\lambda/1000$ Shack-Hartmann wave front sensors (WFS) sampling wave-front deviations at different positions of the FPA. These are similar to \textit{Gaia} WFS \citep{doi:10.1117/12.825240}, and can operate as triggers to ensure optical surface stability before an observation starts. They will also be used as a calibration source, and will enable a deterministic focusing of the \textit{Theia} telescope along the mission. 

Tiling the FPA with CMOS detectors would be ideal to observe some of the brightest possible exoplanet hosts. However, CCDs systematics are much better understood than CMOS, and currently no european CMOS detectors satisfy ESA TRL requirements. Even if CCDs may not be able to readout some of the brightest targets at a high enough rate, there are two possible solutions for a CCD-based FPA. Either two of the detectors in the FPA can be replaced by CMOSes, and thus individual windows could be read at high enough rates, or two of the CCD detectors could be covered with a filter to prevent saturation. During the exoplanet observation, the target star would always need to be located at one of these detectors. To read out the CCD detectors while observing faint targets for DM-related science cases, a shutter mechanism using a slow leaf like the Euclid design can be adopted \cite{2010SPIE.7739E..3KG}. We note that these solutions are only necessary if CCD detectors are chosen over CMOS detectors during the mission Phase-A studies.

Theia most stringent science requirement results in a centroid error calibration of $10^{-5}$ pixel. To monitor the mosaic geometry and the individual quantum efficiency of each pixel and thus to allow the associated systematic errors to be corrected at that level, Theia includes a focal plane metrology subsystem \citep{2016A&A...595A.108C}. This relies on metrology laser feed optical fibers that illuminate the focal plane and form Young's fringes on top of the FPA detectors. These fringes allow to solve the position of each detector and pixel. To measure the inter-pixel and intra-pixel quantum efficiencies, the light phase is modulated and by measuring the fringes at the sub-nanometer level using the information from all the pixels it is possible to determine the QE map and solve the position of reference stars compared to the central target with a differential accuracy better than $\sim1\mu$as. 

%The control of the stability of the FPA and telescope will be ensured by two metrology systems. The focal-plane metrology  will enable to monitor the position of the detectors and pixels relative to each other which is necessary to achieve high precision astrometric measurements. 
%It will also be used to calibrate the inter- and intra-pixel response of the detector during periodic focal plane-calibration. By reaching a calibration of $6\times10^{-5}$ the pixel size, the NEAT-demo\cite{2014SPIE.9143E..4SC} testbed (IPAG/Grenoble)  \cite{2016arXiv160800360C, 2016arXiv160902477C} showed that current technology is mature enough to reach the required level of precision.

In addition to measuring the FPA, the geometry of the structure of the telescope needs monitoring to control time-variable aberrations at sub $\mu$as level. The telescope metrology subsystem is based on linear displacement interferometers. Between each pair of mirror there are six independent interferometric baselines, which are organized into hexapods, each baseline measuring a distance variation. Based on the independent distance determinations, the relative positions and angles between each pair of mirrors can be rigidly determined, as modifications of the geometry impacts all six baselines. Individual baselines capable of reaching 50 picometers enable Theia to attain a soil of systematics control of 125 nas over its FPA.

%The telescope metrology subsystem has three main components: the laser source, the micro-interferometers and the associated electronics.  These laser sources will inject the beam into an optical switcher connected to mono-mode fibers that drive the beam to each micro-interferometer baseline. All have dual redundancy to minimise impact in case of failure.  Each micro-interferometer baseline consists in the micro-interferometer optical bench and an associated retro-reflector. The retro-reflector is a classical corner cube produced from Zerodur \cite{doi:10.1117/12.2055086}. The optical benches consist on a Zerodur bench and associated optics.  Molecular adhesion will be used to fix the optics, the bench and the corner cubes directly to the borders of the \textit{Theia} telescope mirrors.

Finally, although mainly conceived and designed to solve Dark Matter, Habitable ExoEarths and Compact Objects questions, Theia concept is a  truly a general-purpose astrometric infrastructure for the astronomical community during the 30's. 

%About 2 -3 terms per page should be marked for the index.  For this, use the {\bf \textbackslash hbindex\{term\}} command. This command highlights the terms that are being indexed. Please do \emph{not use} the normal \textbackslash index\{\} command, since editors need to be able to locate indexed entries in the compiled document.  Here is an example of an \hbindex{indexed term}.

\begin{acknowledgement}
The authors thank everyone involved in the various astrometric concepts discussed in this chapter for their great efforts. MJ acknowledges support from the Knut and Alice Wallenberg foundation.
\end{acknowledgement}

%  IF you do NOT use bibtex, put comments before the following 2 lines
\bibliographystyle{spbasicHBexo}  %for bibtex
\bibliography{HBexoTemplateBib} %for bibtex-example

\end{document}